\documentclass[twocolumn,american,apl, PRB, notitlepage, groupedaddress, superscriptaddress, twocolumn]{revtex4-2}

\usepackage[T1]{fontenc}
\usepackage[latin9]{inputenc}
\usepackage{babel}
\usepackage{amstext}
\usepackage{amssymb}
\usepackage{graphicx}
\usepackage{esint}
\usepackage{subscript}
\usepackage[unicode=true,pdfusetitle,
 bookmarks=true,bookmarksnumbered=false,bookmarksopen=false,
 breaklinks=false,pdfborder={0 0 1},backref=false,colorlinks=false]
 {hyperref}

\makeatletter
\usepackage{babel}
\usepackage{hyperref}
\date{\today}

\makeatother

\begin{document}
\title{Tunneling study in granular aluminum near the Mott metal-to-insulator
transition}
\author{Aviv Glezer Moshe }
\email{avivmoshe@mail.tau.ac.il}

\affiliation{Raymond and Beverly Sackler School of Physics and Astronomy, Tel Aviv
University, Tel Aviv, Israel}
\affiliation{Department of Physics and Department of Electrical and Electronic
Engineering, Ariel University, P.O.B. 3, Ariel 40700}
\author{Gal Tuvia}
\affiliation{Raymond and Beverly Sackler School of Physics and Astronomy, Tel Aviv
University, Tel Aviv, Israel}
\author{Shilo Avraham}
\affiliation{Raymond and Beverly Sackler School of Physics and Astronomy, Tel Aviv
University, Tel Aviv, Israel}
\affiliation{Department of Physics and Department of Electrical and Electronic
Engineering, Ariel University, P.O.B. 3, Ariel 40700}
\author{Eli Farber}
\affiliation{Department of Physics and Department of Electrical and Electronic
Engineering, Ariel University, P.O.B. 3, Ariel 40700}
\author{Guy Deutscher}
\affiliation{Raymond and Beverly Sackler School of Physics and Astronomy, Tel Aviv
University, Tel Aviv, Israel}
\begin{abstract}
We find excellent agreement between tunneling and optical conductivity
gap values in superconducting granular aluminum films, up to the metal-to-insulator
transition. This behavior, in strong contrast with that recently reported
in atomically disordered samples for which the optical gap becomes
smaller than the tunneling gap, suggests that disorder is not at the
origin of the transition. The large increase seen in the strong coupling
ratio and a finite value of the gap at $T_{\text{c}}$ near the metal-to-insulator
transition are consistent with a BCS to BEC crossover.
\end{abstract}
\maketitle

\section{Introduction}

In conventional BCS theory \citep{Bardeen1957} the superconducting
gap $\Delta$ in the density of states (DOS) is equal to half of the
absorption threshold for photons $2\Omega$. This statement remains
correct for the case of a mildly disordered dirty superconductor (SC)
\citep{Anderson1959}, i.e. $E_{F}/\Delta\gg k_{F}l\gg1$ where $E_{F}$
is the Fermi energy, $k_{F}$ is the Fermi wave vector and $l$ is
the mean free path, far from a disorder driven Anderson metal-to-insulator
(M/I) transition. In that case, the BCS based Mattis-Bardeen (MB)
theory \citep{Mattis1958} describes the frequency dependent complex
optical conductivity $\sigma(\omega)=\sigma_{1}+i\sigma_{2}$ exceptionally
well. Its real part drops to zero below $\hbar\omega=2\Delta$. It
has a minimum at that frequency at finite temperatures.

When disorder on the atomic scale is increased, Larkin and Ovchinnikov
(LO) theory predicts that the coherence peak in the DOS located at
$\Delta$ broadens. It has a universal shape which depends on a single
parameter $\eta$ that increases with disorder. As it does, the absorption
threshold becomes smaller than the mean value of the gap $\Delta$
given approximately by the location of the maximum in the DOS \citep{larkin1972density}.
The DOS will differ from the BCS prediction and the optical conductivity
will not follow the MB theory. The energy scale $\Omega$ as determined
by optical conductivity will become smaller than the energy scale
$\Delta$ as determined by tunneling. In the case that the system
is composed of metallic grains weakly coupled together, disorder consists
of the broad distribution of the grain size and the coupling between
the grains. In that case as well, the gap will be inhomogeneous resulting
in similar differences between optical and tunneling gap measurements.

If one assumes that the system is homogeneous, the gap should be homogeneous
as well. Determinations of the gap by optical and tunneling measurements
should give the same results. The insulating state can be driven by
the competition between the bandwidth and the Coulomb interaction
i.e., a M/I transition of the Mott type. 

Experiments on atomically disordered SCs such as InO\textsubscript{x}
and NbN\textsubscript{x} show indeed that as disorder increases,
$\Omega$ indeed becomes smaller than $\Delta$ \citep{Sherman2015,Cheng2016}.
In the case of NbN\textsubscript{x}, the experiment is consistent
with the LO theory \citep{Cheng2016}. 

In this work we present a study of granular aluminum (grAl) films
approaching the M/I transition using planar tunneling junctions. We
find that in these films, consisting of nano-scale metallic grains
weakly coupled together, the tunneling energy scale $\Delta$ and
the optical energy scale $\Omega$ \citep{Moshe2019} remain equal
to each other up to the M/I transition. This implies that disorder
plays only a minor role in the M/I transition of grAl, contrary to
the case of NbN\textsubscript{x} and InO\textsubscript{x}. We ascribe
this different behavior to a M/I transition, triggered by the electrostatic
charging energy of the small grains. In that case the M/I transition
is expected to be of the Mott type \citep{Bachar2015,Moshe2019,GlezerMoshe2020,GlezerMoshe2020a}
rather than an Anderson type as seen in atomically disordered films.
According to dynamical mean field theory (DMFT) calculations, in a
Mott transition a narrow region of finite and unchanged DOS relatively
to the metallic state is formed at the Fermi level \citep{Georges1996}.
Hence, a BCS-BEC crossover is expected as the width of this region
approaches the superconducting gap. In that scenario there is no known
reason why the energy scales $\Delta$ and $\Omega$ should become
different.

Rather, in the smooth crossover from highly overlapping fermionic
pairs ($k_{F}\xi_{\text{pair}}\gg1$) to tightly bound bosonic molecules
($k_{F}\xi_{\text{pair}}\ll1$), it is the energy scales $\Delta$
(the pairing energy) and $T_{\text{c}}$ (the condensation temperature)
that are progressively decoupled from each other \citep{Strinati2018}.
In the BCS side they are linked by the weak-coupling relation $2\Delta=3.53k_{\text{B}}T_{\text{c}}$
while in the so called unitary limit where $k_{F}\xi_{\text{pair}}\sim1$
(near the BEC side) this ratio is increasing up to about 6 \citep{Pisani2018,Pisani2018a}.
It is precisely this behavior that we have observed.

\smallskip{}

\section{Methods}

The grAl films were prepared by thermally evaporating clean Al pellets
under controlled O\textsubscript{2} pressure onto liquid nitrogen
cooled substrate or onto room temperature warm substrate, for more
details see \citep{Moshe2019}. The role that the substrate temperature
plays is crucial, since it determines the grain size, its distribution
and the maximum $T_{c}$. For substrates held at room temperature
the maximum $T_{c}$ is about 2.2 K \citep{Deutscher1973,Dynes1984,Levy-Bertrand2019}
and the grain size is about $3\pm1\,$nm \citep{Deutscher1973a} whereas
for substrates held at liquid nitrogen temperature the maximum $T_{c}$
is about 3.2 K and the grain size is about $2\pm0.5\,$nm \citep{Deutscher1973,Lerer2014}.

In order to obtain samples with varying normal state resistivity along
the phase diagram we varied the O\textsubscript{2} pressure while
keeping the deposition rate around $\text{5}\pm\text{1}\,\mathring{\text{A}}\text{/s}$
. To obtain high quality, homogeneous samples (in the sense of grain
coupling), we made sure that the pressure and deposition rate were
stable during deposition.

For the tunneling experiments we first evaporate 50 nm of grAl film
onto liquid nitrogen cooled Al\textsubscript{2}O\textsubscript{3}
substrate, patterned to a rectangle with dimensions of 0.5x4 mm, by
using a hard mask. After evaporation, we expose the film to atmosphere
for a few hours, resulting in a native thin oxide barrier. We then
use a second mask to evaporate two 50 nm thick, 0.5 mm wide aluminum
strips, perpendicular to the grAl film, resulting in a grAl-oxide-Al
tunneling junctions. In addition, this geometry allows us to measure
the resistivity of the film vs. temperature and therefore to determine
the critical temperature $T_{c}$ and the normal state resistivity
$\rho_{4.2\text{K}}$ for each sample.

For each junction we took current-voltage (I vs. V) and differential
conductance-voltage (dI/dV vs. V) characteristics using Keithley 6221
current source and Keithley 2182A nano-voltmeter, operated in differential
conductance mode. The same set of equipment, operated in delta mode,
was used for the resistivity measurements. Measurements were taken
in three different systems: a commercial \textsuperscript{3}He probe
by Quantum Design PPMS with a base temperature of 0.4 K, commercial
Oxford Instruments Triton 400 dilution refrigerator with a base temperature
of \textasciitilde 20 mK and a liquid \textsuperscript{4}He bath
cryostat with a base temperature of about 1.5 K. 

Each dI/dV spectrum was fitted to a standard tunneling form \citep{Tinkham2004}

\begin{equation}
\begin{array}{c}
\left.\frac{dI}{dV}\right|_{V}\propto\\
\frac{d}{dV}\left\{ \int_{-\infty}^{+\infty}N_{s}(E)N_{n}(E+eV)\{f(E)-f(E+eV)\}dE\right\} 
\end{array}\label{eq:Dynes}
\end{equation}
where $f(E)$ is the Fermi function, $N_{n}(E)$ is the constant density
of states (DOS) of the Al electrode and $N_{s}(E)=\text{Re\ensuremath{\{(E-i\Gamma)/[(E-i\Gamma)^{2}-\Delta^{2}]^{1/2}\}}}$
is the Dynes DOS \citep{Dynes1978} that takes into account the coherence
peaks broadening $\Gamma$. This approach allows us to obtain the
superconducting gap $\Delta(T)$ and the broadening parameter $\Gamma(T)$.
For the highest resistivity sample studied we have normalized the
$dI/dV$ data prior to fitting it with Eq. \ref{eq:Dynes} in order
to avoid the influence of the high voltage $dI/dV\propto\sqrt{V}$
behavior \citep{Dynes1981,Altshuler1979} on the superconducting DOS.
For more details on the data analysis, see supplemental material \citep{SM}.

To extend the comparison between tunneling and optical conductivity
to samples having larger grain size, additional samples were prepared
on $10\times10\times2\,$mm\textsuperscript{3} MgO or $10\times10\times1\,$mm\textsuperscript{3}
(LaAlO\textsubscript{3})\textsubscript{0.3}- (SrAl\textsubscript{0.5}Ta\textsubscript{0.5}O\textsubscript{3})\textsubscript{0.7}
(LSAT) substrates, held at room temperature, for THz optical spectroscopy.
All samples are 100 nm thick, except the lowest resistivity one, which
is 40 nm thick. Measurements were done by utilizing a quasi-optical
Mach-Zehnder interferometer with backward-wave oscillators (BWO) in
a frequency range of 3-17 cm\textsuperscript{-1} (or about 0.1-0.5
THz). Commercial optical \textsuperscript{4}He cryostat with a home
built sample holder, provides us dynamic temperature range down to
1.5 K. The complex transmission was measured for all samples at 4.2
K and at various temperatures close to and below $T_{c}$ down to
1.5 K. We then obtain the optical conductivity $\sigma(\omega)=\sigma_{1}(\omega)+i\sigma_{2}(\omega)$
from the measured complex transmission function. We take $\sigma_{1}(\omega)$
at 4.2 K as the normal-state conductivity and then fit the resulted
$\sigma(\omega,T)/\sigma_{n}$ to Mattis-Bardeen (MB) theory \citep{Mattis1958},
as appropriate for a dirty limit SC \citep{Pracht2016,Moshe2019}.
The only fitting parameter is $\Omega(T)$ which we fit to the BCS
gap equation \citep{Bardeen1957} in order to obtain $\Omega(T\rightarrow0)=\Omega_{0}$.
For more details, see supplemental material \citep{SM}. We used Dynes
et al. \citep{Dynes1984} tunneling data for samples prepared under
similar conditions, in order to compare between the optical and tunneling
gaps of the larger grain size samples.

\smallskip{}

\section{Results}

The resistivity curves for selected 2 nm grain samples are shown in
Fig. \ref{fig:RT curves}. We take $T_{\text{c}}$ as the temperature
where the resistivity falls to one percent of its normal state value.
Tunneling differential conductance data for samples having a resistivity
value up to \textasciitilde 1500 $\mu\Omega\,\text{cm}$ was taken
from $T_{\text{c}}$ down to \textasciitilde 1.6-1.8 K, above the
critical temperature of the Al counter-electrode. We note that for
a few measured junctions $\Gamma$ was found to be \textasciitilde 0.1
meV and roughly temperature independent, in contrast to all other
samples showing a decreasing $\Gamma$ with temperature (see for example
Figs. S4 and S5 in the supplemental material \citep{SM}). We find
this as an indication that for these samples the barrier quality is
poor and therefore we have discarded their $\Delta(0)$ values. We
note that for samples where $\Gamma>\Delta$ near and below $T_{\text{c}}$
(see Fig. \ref{fig:highrho_nearTc} and Fig. S4) the reliability of
the obtained values from the fit is low.

The zero temperature value of the gap was determined by fitting $\Delta(T)$
to the BCS gap equation \citep{Bardeen1957}. For samples having very
high resistivity values approaching the Metal to Insulator transition,
and substantially reduced critical temperatures, tunneling data was
also taken at lower temperatures down to 30 mK. In that case data
was taken both in zero field, and in a magnetic field of \textasciitilde 0.1
T, high enough to quench superconductivity in the Al counter-electrode. 
\begin{center}
\begin{figure}
\centering{}\includegraphics[width=1\columnwidth]{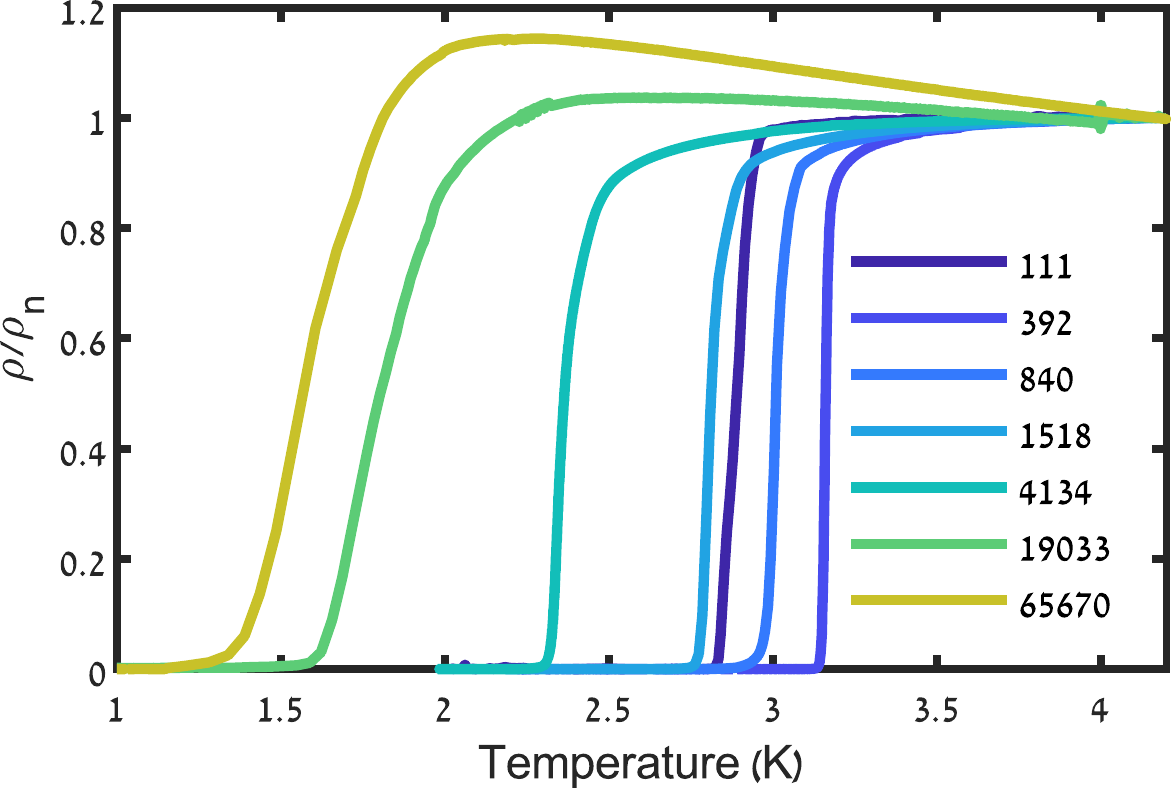}\caption{Normalized resistivity curves for selected samples along the phase
diagram. The legend corresponds to the normal state resistivity value
$\rho_{n}$, expressed in $\mu\Omega\,\text{cm}$. Note how the transition
remains relatively sharp, even for very high resistivity values, for
which superconductivity is destroyed in atomically disordered samples
such as NbN\protect\textsubscript{x} \citep{Chand2012,Mondal2011,Mondal2011a}.
\label{fig:RT curves}}
\end{figure}
\par\end{center}

We observed a few percent change in the gap relatively to its zero
field extrapolated value up to a field of 1 Tesla. The change in the
value of $\Gamma$ with increasing magnetic field is more significant,
as expected since it introduces spatial inhomogeneities in the gap
due to the normal cores of about 6-7 nm in size \citep{GlezerMoshe2020a}
(see supplemental material for more details \citep{SM}). The orbital
upper critical field is estimated as about 4 T as measured in a previous
work on similar samples \citep{GlezerMoshe2020a}. We therefore conclude
that the applied field of \textasciitilde 0.1 T does not significantly
affect the measured DOS of the grAl film. We show in Fig. \ref{Field_4100}
the tunneling data for a sample having a resistivity value of \textasciitilde 4,000
$\mu\Omega\,\text{cm}$ measured at 30 mK, both in zero applied field
and 0.1 T applied field. Both sets have been fitted to a Dynes form,
in zero applied field taking for the Al counter-electrode a gap of
0.24 meV and for the grAl film a gap of 0.394 meV extrapolated down
to zero field (full field dependent analysis can be found in the supplemental
material \citep{SM}). A similar analysis was performed on a \textasciitilde 66,000
$\mu\Omega\,\text{cm}$ sample at 0.4 K, resulting in a gap of 0.26
meV for the Al counter electrode with the grAl film having a gap of
0.32 meV, even though its critical temperature of 1.37 K is lower
than the one measured for the Al counter electrode of 1.78 K. More
details can be found in the supplemental material \citep{SM}.

Tunneling gap values $\Delta_{0}$ for 2 nm grain size samples having
resistivity values ranging from \textasciitilde 100 to \textasciitilde 66,000
$\mu\Omega\,\text{cm}$ are shown in Fig. \ref{fig:PD_Tc_del_coup_2nm}
where they are compared to optical gap values $\Omega_{0}$ for samples
having resistivity values up to \textasciitilde 8000 $\mu\Omega\,\text{cm}$.
It can be clearly seen that they both sets are in excellent agreement.
This point is further illustrated in Fig. \ref{fig:dIdV_THz_Tunn},
where $\Delta$ as indicated by the coherence peak position for the
tunneling data and $\Omega$ by the minimum in $\sigma_{1,s}$ for
the optical data are seen to coincide. We note that the decrease of
$\Delta_{0}$ at high resistivity values is substantially weaker than
that of the critical temperature. As shown in Fig. \ref{fig:PD_Tc_del_coup_2nm},
the trend of increasing strong coupling ratio previously reported
for the optical gap \citep{Pracht2016,Moshe2019} can be clearly seen
here. It reaches a value of 5.64 for the \textasciitilde 66,000 $\mu\Omega\,$cm
sample.
\begin{center}
\begin{figure}
\centering{}\includegraphics[width=1\columnwidth]{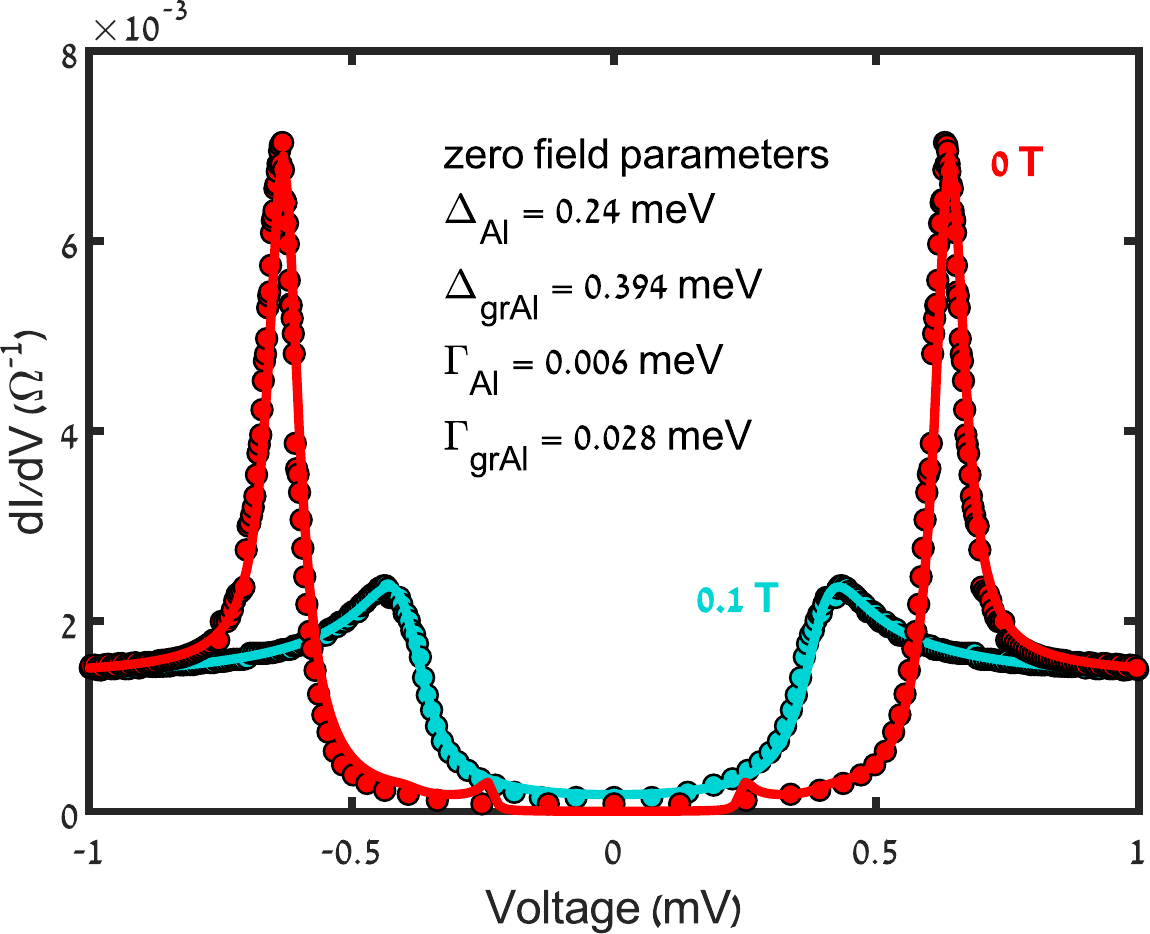}\caption{Tunneling differential conductances in zero field (SIS) and 0.1 T
field (SIN) for a sample having a resistivity of 4134 $\mu\Omega\,\text{cm}$,
measured at 30 mK. Circles are data points and lines are fits to Eq.
\ref{eq:Dynes}. Extended data in magnetic fields up to 1 T can be
found in the supplemental material \citep{SM}.}
\label{Field_4100}
\end{figure}
\par\end{center}

\begin{figure}
\centering{}\includegraphics[width=1\columnwidth]{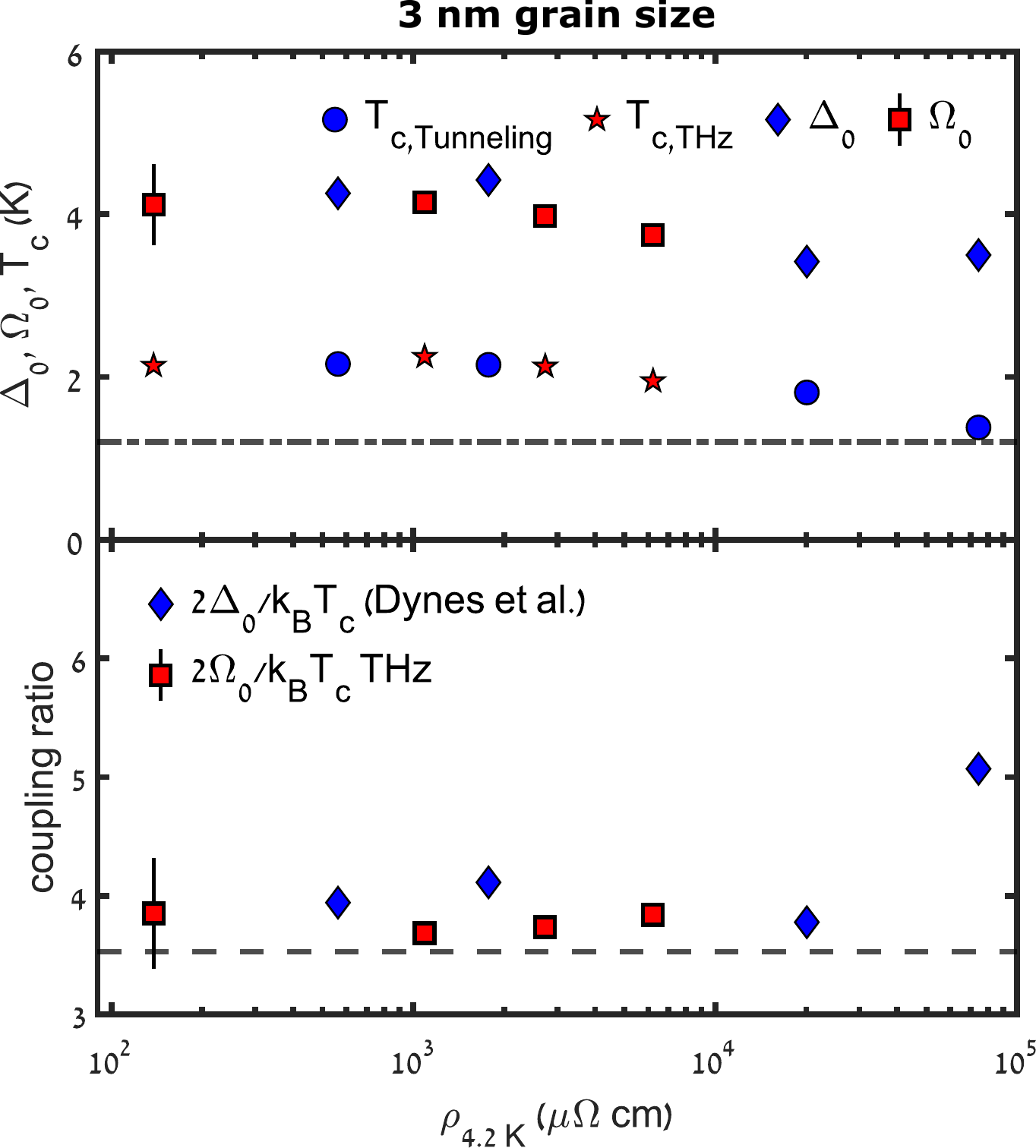}\caption{$T_{c}$ , $\Delta_{0}$, $\Omega_{0}$ and coupling ratio vs. $\rho_{n}=\rho_{4.2\,\text{K}}$
for samples having a grain size of 3 nm. Top: $T_{c}$ , $\Delta_{0}$
and $\Omega_{0}$ vs. $\rho_{n}$. $\Omega_{0}$ is from THz spectroscopy
data (this work) and $\Delta_{0}$ is from previous tunneling data
(Dynes \textit{et al}. \citep{Dynes1984}). The semi-dashed line marks
the $T_{c}$ of pure Al, 1.2 K. The error bars on $\Omega_{0}$ reflect
the uncertainty in the MB fit of $\sigma_{1}(\omega)$. The error
bars on $\Delta_{0}$ (smaller than the symbol size) reflect the uncertainty
in the Dynes fit of the differential conductance. Bottom: coupling
ratio vs. $\rho_{n}$. $2\Omega_{0}/k_{\text{B}}T_{\text{c}}$ has
been obtained from this work THz spectroscopy data and $2\Delta_{0}/k_{\text{B}}T_{\text{c}}$
is taken from a previous tunneling work by Dynes \textit{et al.} \citep{Dynes1984}.
The dashed line marks the BCS weak coupling ratio, 3.53. The error
bars reflect the uncertainty in $\Omega_{0}$ and $\Delta_{0}$. \label{fig:PD_Tc_del_coup_3nm}}
\end{figure}

We can gain more insight on the role of the grain size by making a
similar comparison between optical and tunneling data for samples
with a grain size of 3 nm. We have compared $\Omega_{0}$ as measured
from the optical conductivity data (see supplemental material \citep{SM})
to the tunneling $\Delta_{0}$ as measured by Dynes \textit{et al.}
\citep{Dynes1984}\textit{, }see top panel of Fig. \ref{fig:PD_Tc_del_coup_3nm}.
As for the case of 2 nm grain size samples, there is good agreement
between the tunneling and optical conductivity gaps. As well, the
variation of $T_{c}$ is in line with the tunneling data. However,
we see that the coupling ratio (Fig. \ref{fig:PD_Tc_del_coup_3nm},
bottom panel) does not change as rapidly in comparison to the 2 nm
grain size samples. It does however reach a value close to 5 for the
highest resistivity sample.
\begin{center}
\begin{figure}
\centering{}\includegraphics[width=1\columnwidth]{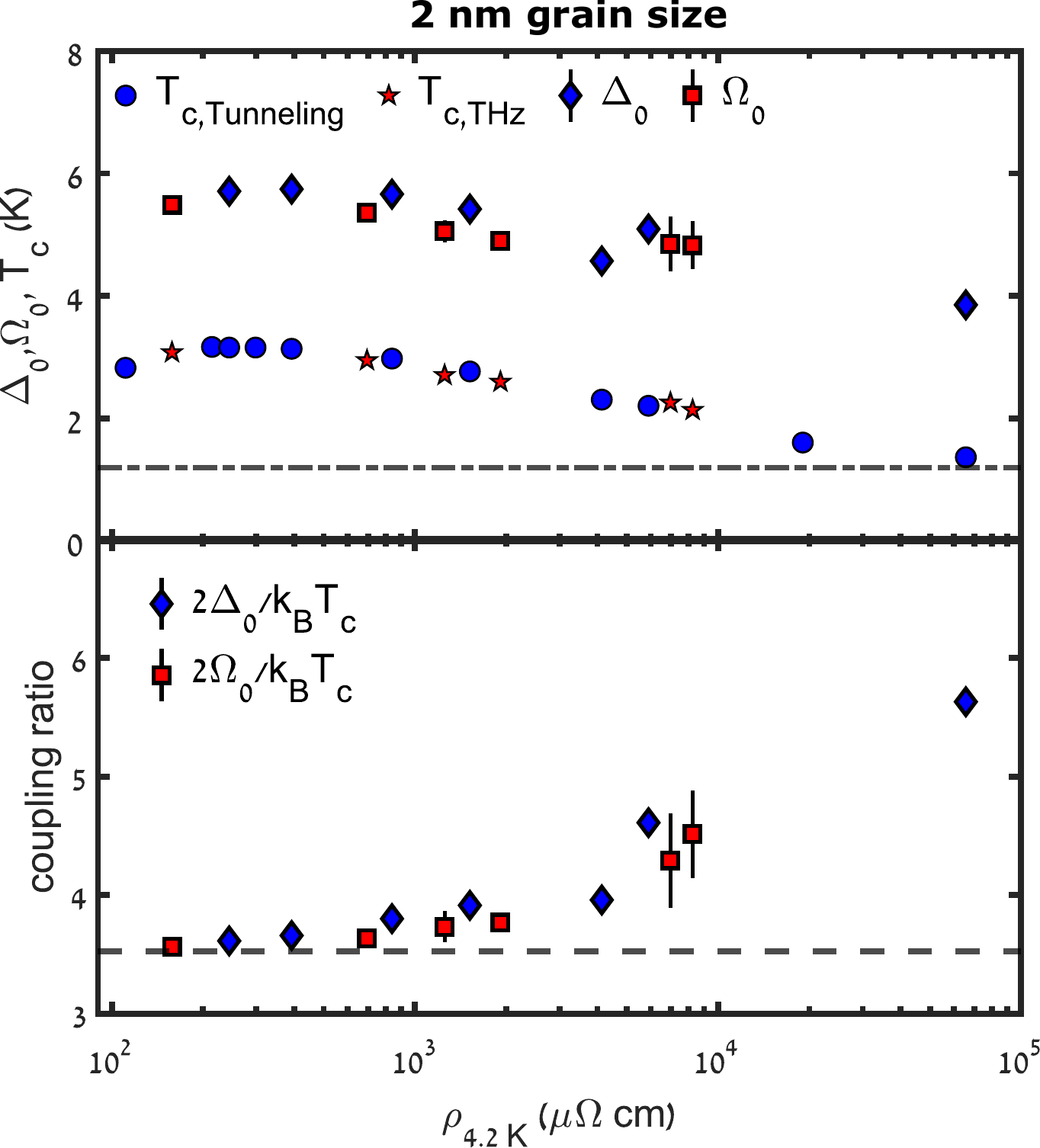}\caption{$T_{c}$ , $\Delta_{0}$, $\Omega_{0}$ and coupling ratio vs. $\rho_{n}=\rho_{4.2\,\text{K}}$
for samples having a grain size of 2 nm. Top: $T_{c}$ , $\Delta_{0}$
and $\Omega_{0}$ vs. $\rho_{n}$. $\Omega_{0}$ is from our previous
THz spectroscopy data \citep{Moshe2019} and $\Delta_{0}$ is from
the tunneling data (this work). Note that for the sample with a resistivity
value close to 20,000 $\mu\Omega$ cm, we only measured $T_{c}$ and
for few of the low $\rho$ samples we only show $T_{c}$, due to poor
junction quality. The semi-dashed line marks the $T_{c}$ of pure
Al, 1.2 K. The error bars on $\Omega_{0}$ reflect the uncertainty
in the MB fit of $\sigma_{1}(\omega)$. The error bars on $\Delta_{0}$
(smaller than the symbol size) reflect the uncertainty in the Dynes
fit of the differential conductance. Bottom: coupling ratio vs. $\rho_{n}$.
The values of $2\Omega_{0}/k_{\text{B}}T_{\text{c}}$ and $2\Delta_{0}/k_{\text{B}}T_{\text{c}}$
have been obtained from our previous THz spectroscopy data \citep{Moshe2019}
and from this work tunneling data. The dashed line marks the BCS weak
coupling ratio, 3.53. The error bars reflect the uncertainty in $\Omega_{0}$
and $\Delta_{0}$. \label{fig:PD_Tc_del_coup_2nm}}
\end{figure}
\par\end{center}

\begin{center}
\begin{figure}
\begin{centering}
\includegraphics[width=1\columnwidth]{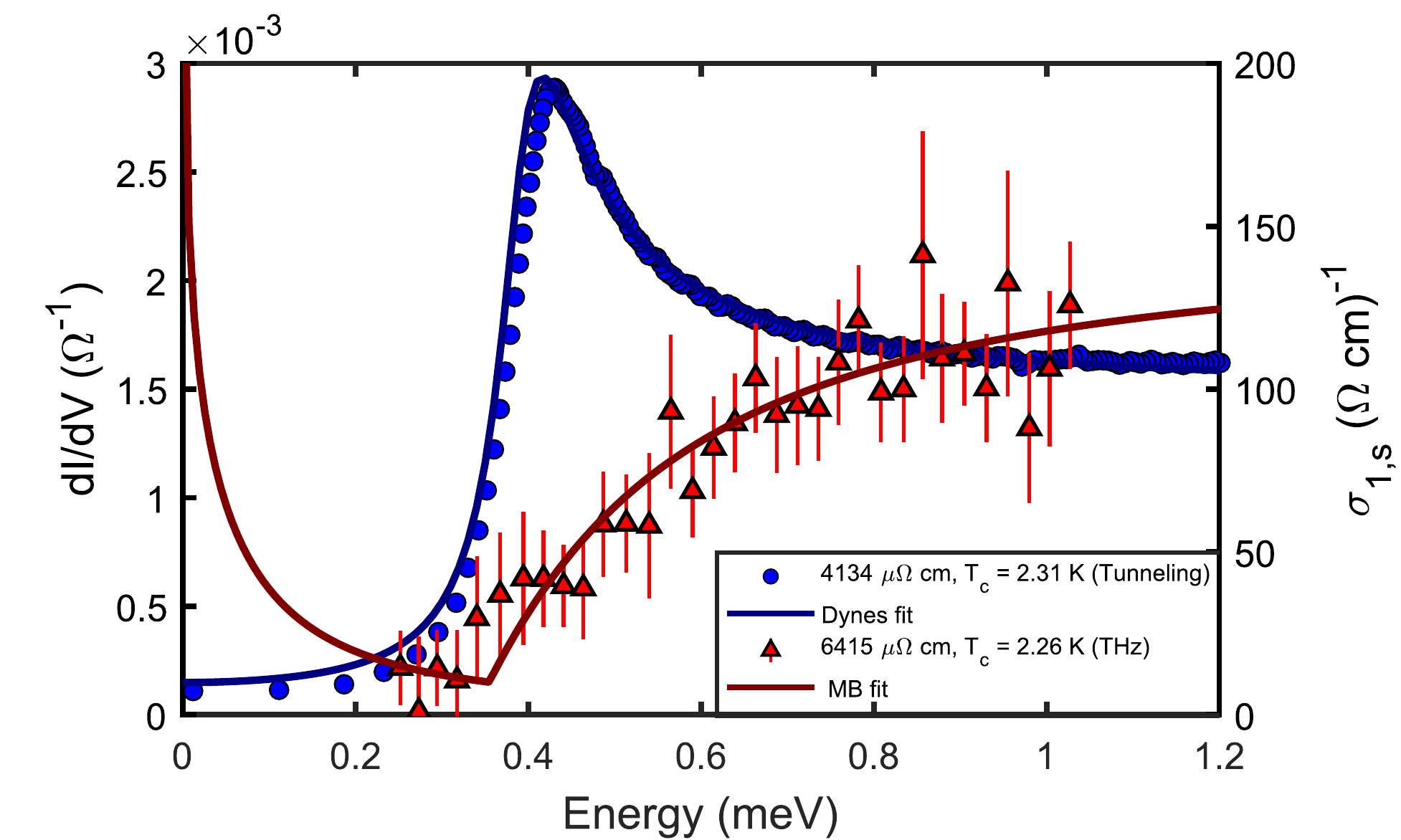} 
\par\end{centering}
\caption{Illustration of the agreement of $\Delta$ and $\Omega$ in samples
(with grain size of 2 nm) with relatively close values of $\rho$
and $T_{c}$. The left axis corresponds to the differential conductance
$dI/dV$, taken at 24 mK, along a fit to a Dynes form. The right axis
corresponds to the real part of the optical conductivity $\sigma_{1,s}$,
taken at 1.5 K, along a fit to MB theory. Note that for the THz data
the horizontal axis has been divided by a factor of two to take into
account that the absorption threshold corresponds to $2\Omega$. The
error bars on $\sigma_{1,s}$ reflect the experimental accuracy in
the optical transmission mainly due to standing waves in the optical
cryostat. The main purpose of this figure is to show that tunneling
and optical conductivity give basically the same gap values. See Glezer
Moshe \textit{et al}. \citep{Moshe2019} detailed discussion of the
optical conductivity in the 2 nm grain size samples. \label{fig:dIdV_THz_Tunn}}
\end{figure}
\par\end{center}

Besides the anomalously large values of the strong coupling ratio
seen in high resistivity samples, a closer look at their tunneling
data near $T_{c}$ reveals that their gap is not being fully closed,
as it should in a conventional SC. We have observed such behavior
for resistivity values above \textasciitilde 800 $\mu\Omega$ cm,
within a temperature range which increases with sample resistivity
(see supplemental Fig. S4 \citep{SM}). This behavior is well demonstrated
in Fig. \ref{fig:highrho_nearTc} for the \textasciitilde 4,000 $\mu\Omega$
cm sample. Quite near $T_{\text{c}}$ its gap is still around 0.3
meV, while at low temperature it reaches 0.394 meV, see Fig. \ref{Field_4100}.

\begin{figure}
\centering{}\includegraphics[width=1\columnwidth]{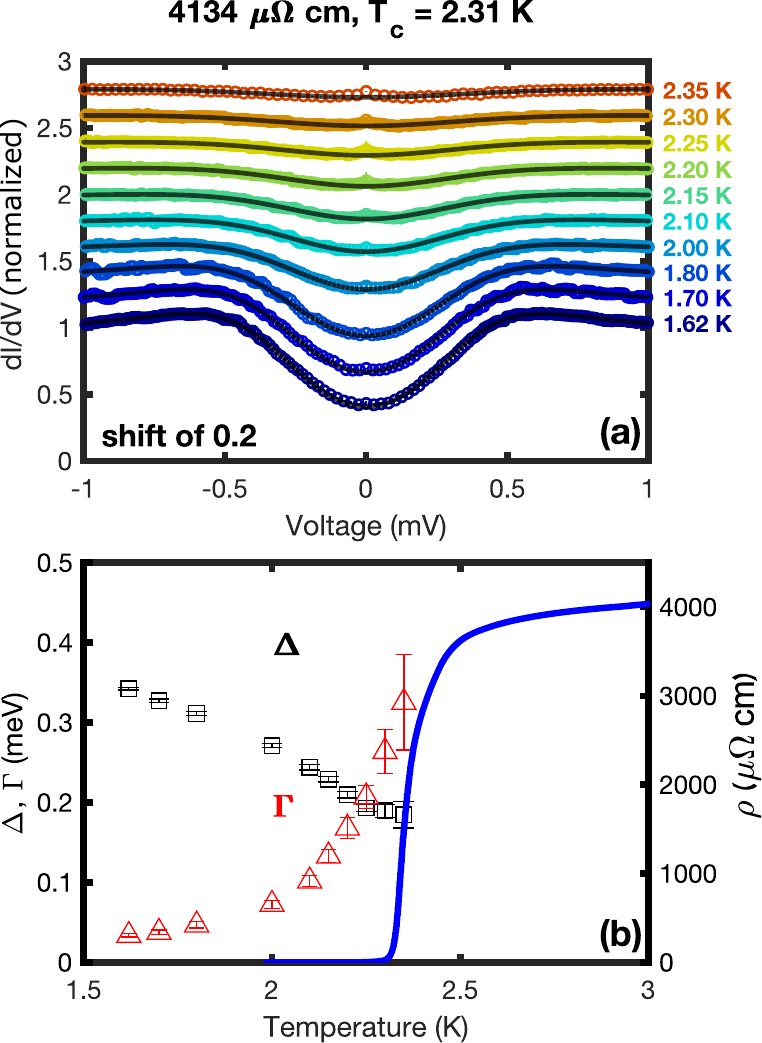}\caption{High resistivity data near and below the critical temperature. (a)
Normalized differential conductance vs. voltage, taken close to the
critical temperature, along a fit to a Dynes form. The data is shifted
for clarity. Details on the normalization procedure can be found in
the supplemental material \citep{SM}. (b) $\Delta$ and $\Gamma$
(left axis in units of meV) vs. temperature, as obtained from the
fits, along the resistivity $\rho$ (right axis). Note that $\Delta$
remains finite near $T_{c}$, as predicted in a BCS-BEC crossover
scenario near unitarity \citep{Pisani2018,Pisani2018a}. The data
was taken using a \protect\textsuperscript{4}He bath cryostat, prior
to the dilution refrigerator measurements (shown in Fig. \ref{fig:dIdV_THz_Tunn}).
Note that since the sample sheet resistance was comparable to the
junction resistance, we could not take reliable data above $T_{c}$.
The error bars on $\Delta$ and $\Gamma$ reflect the uncertainty
in the Dynes fit of the differential conductance. \label{fig:highrho_nearTc}}
\end{figure}

\smallskip{}

\section{Discussion}

In atomically disordered SC, the M/I transition is of the Anderson
type. As disorder increases, the tunneling and optical conductivity
gaps are being progressively separated, as was observed in NbN\textsubscript{x}
\citep{Chand2012,Mondal2011a,Mondal2011,Cheng2016,Sherman2015} and
InO\textsubscript{x} \citep{Sherman2014,Sherman2015}. The separation
can be described quite well by the theory of LO \citep{larkin1972density}
which predicts that $\Omega=(1-\eta^{2/3})^{3/2}\Delta$, where $\eta$
is a depairing parameter that increases with disorder. This theory
shows good agreement with experiment \citep{Cheng2016}. This is in
strike contrast to what we observe in grAl, namely there is a good
agreement between the tunneling and optical conductivity gaps up to
very high resistivity values, much higher than observed in NbN\textsubscript{x}.
The work of Feigel\textquoteright man and Skvortsov \citep{Feigelman2012}
confirms that strong disorder should manifest itself as a different
tunneling and optical gaps. They predict that in that case the parameter
$\Gamma$ should not be small compared to the gap. We have compared
values of $\Gamma$ to the differential conductance peak position
and found it to be small even up to resistivities of several 1,000
$\mu\Omega$ cm. Therefore, disorder effects are small in our case
and plays little role, if any, in the M/I transition of grAl. Recent
scanning tunneling microscope (STM) measurements on samples prepared
on room temperature substrates shows that the gap value does not change
within the sample \citep{Yang2020}, a clear sign for low disorder. 

Rather our two central results: the large increase of the strong coupling
ratio and the anomalous behavior of the gap near $T_{\text{c}}$,
point out to a different interpretation: a BCS-BEC cross-over as the
M/I transition is approached. We find remarkable that for our highest
resistivity 2 nm grain size sample, which is very close to being insulating,
the value close to 6 that we have obtained for the strong coupling
ratio is that predicted at cross-over \citep{Pisani2018,Pisani2018a}.
We also note that a similar value (about 5) is obtained from tunneling
data of Dynes \textit{et al.} for a 3 nm grain size sample having
a very close resistivity value \citep{Dynes1984}. A non-closure of
the gap at $T_{\text{c}}$, as can be seen in Fig. \ref{fig:highrho_nearTc},
is also expected at a BCS to BEC cross-over. This is because of the
different nature of excitations leading to the destruction of superconductivity
in the BCS and BEC regimes: pair breaking in the former, collective
excitations in the latter. At cross-over pair formation already occurs
above $T_{\text{c}}$.

A Mott transition has been shown to occur in granular aluminum films
having a resistivity of about 50,000 $\mu\Omega$ cm at room temperature
\citep{Bachar2015}, very close to that of our highest resistivity
sample where the largest strong coupling ratio is seen (having a room
temperature resistivity of about 20,000 $\mu\Omega$ cm). A Mott driven
BCS to BEC crossover thus provides an explanation for our observations.
It can be expected that the split between a lower and a higher Hubbard
bands with a narrow DOS peak subsisting in the middle, predicted by
DMFT theory \citep{Georges1996}, should impact the tunneling conductance
near a Mott transition. However, the absence of a tunneling theory
for that regime does not allow at this point a quantitative analysis
of the tunneling data. In addition, recent calculations of a model
that takes into account both Mott behavior and SC shows as well that
the coupling ratio exceeds that of the BCS weak coupling value \citep{Phillips2020}.
Another piece of evidence which supports this picture is the fact
that the coherence length $\xi$ decreases with resistivity down until
it saturates to a value of about 6-7 nm (as obtained from upper critical
field measurements \citep{GlezerMoshe2020a}). This saturation of
the coherence length is a characteristic of the BCS-BEC crossover
regime \citep{Spuntarelli2010}.

In conclusion, we have measured the evolution of the SC gap $\Delta$
by tunneling and THz spectroscopy in granular aluminum films, and
have found them to be in good agreement up to the vicinity of the
M/I transition. We have shown that the coupling ratio $2\Delta/k_{B}T_{c}$
is increasing up to about 6 for 2 nm grain size samples, and that
as the transition is approached there is no closure of the gap at
$T_{\text{c}}$. Both features are characteristic of a BCS-BCS crossover,
which we have proposed is induced by the vicinity to a Mott transition.
A quantitative analysis of the tunneling data, in particular regarding
the behavior of the conductance near $T_{\text{c}}$ for high resistivity
samples, is needed to bring a definite proof of a BCS-BEC crossover.
This requires a tunneling theory for that regime that is not yet available.

\smallskip{}

\section*{Acknowledgments}

Helpful discussions with Ioan Pop and Giancarlo Strinati are gratefully
acknowledged. We thank Yoram Dagan for granting us access to his dilution
refrigerator.

\newpage

\bibliographystyle{apsrev4-2}

\end{document}